\documentclass[aps,pre,showpacs,twocolumn]{revtex4}
\usepackage{bm}
\usepackage{graphicx}
\bibstyle{approve.bib}
\usepackage{amssymb}
\usepackage{amsmath}
\usepackage{esint}
\usepackage{epstopdf}
\usepackage{color}
\usepackage{gensymb}
\usepackage{mathtools}
\usepackage{suffix}
\usepackage{fancyhdr}

\newcommand{\be}{\begin{equation}}
\newcommand{\ee}{\end{equation}}
\newcommand{\bea}{\begin{eqnarray}}
\newcommand{\eea}{\end{eqnarray}}

\newcommand{\br}{\mathbf{r}}

\newcommand{\bE}{\mathbf{E}}

\newcommand{\lb}{\lambda_b}
\newcommand{\lel}{\lambda_d}

\newcommand{\e}{\varepsilon}

\newcommand{\tG}{\tilde{G}}

\begin{document}

\title{Facilitated polymer capture by charge inverted electroosmotic flow in voltage-driven polymer translocation}

\author{Sahin Buyukdagli$^{1,2}$\footnote{email:~\texttt{buyukdagli@fen.bilkent.edu.tr}}}
\affiliation{$^{1}$Department of Physics, Bilkent University, Ankara 06800, Turkey\\
$^{2}$QTF Centre of Excellence, Department of Applied Physics, Aalto University, FI-00076 Aalto, Finland.}

\begin{abstract}

The optimal functioning of nanopore-based biosensing tools necessitates rapid polymer capture from the ion reservoir. We identify an ionic correlation-induced transport mechanism that provides this condition without the chemical modification of the polymer or the pore surface.  In the typical experimental configuration where a negatively charged silicon-based pore confines a 1:1 electrolyte solution,  anionic polymer capture is limited by electrostatic polymer-membrane repulsion and the electroosmotic (EO) flow. Added multivalent cations suppress the electrostatic barrier and revers the pore charge, inverting the direction of the EO flow that drags the polymer to the trans side. This inverted EO flow can be used to speed up polymer capture from the reservoir and to transport weakly or non-uniformly charged polymers that cannot be controlled by electrophoresis.

\end{abstract}


\date{\today}
\maketitle

\section{Introduction}

Bionanotechnology occupies a central position among the emerging scientific disciplines of the twenty-first century. This fast-growing field offers various bioanalytical strategies that make use of nanoscale physical phenomena~\cite{rev1,rev2}. Among these techniques, polymer translocation has been a major focus during the last two decades~\cite{Tapsarev}. A typical translocation process consists of guiding a biopolymer through a nanopore and reading its sequence from the ionic current perturbations caused by the molecule~\cite{e1,e2,e3,e5,e6,e10,e13,e19,e21}. By relying mainly on the electrohydrodynamics of the confined polymer-liquid complex, this biosensing method allows to bypass the biochemical modification of the polymer, thereby providing a fast and cheap sequencing of the molecule.  

Due to the working principle of polymer translocation, the predictive design of translocation tools necessitates the through characterization of the entropic, electrostatic, and hydrodynamic effects governing the system.  Entropic effects associated with polymer conformations and hard-core polymer-pore interactions have been intensively addressed by numerical simulations~\cite{n1,n2,n5}. The electrohydrodynamics of polymer translocation has been also considered by mean-field (MF) electrostatic theories~\cite{the2,the3,the4,mut1,mut2,the8,Hatlo,hooger,mut3,mut4,mf} and simulations~\cite{aks1,aks2,aks3,aks4,Luan,HLsim}. 

In certain physiological conditions relevant to polymer translocation, such as strongly charges pores or in the presence of multivalent ions, MF electrostatics fails and charge correlations have to be included. For example, an accurate readout of the ionic current signal is known to require a long enough translocation time~\cite{e6}. Simulations by Luan et al.~\cite{aks3,Luan} and our former theoretical study~\cite{the16} showed that added polyvalent cations can fulfill this condition by cancelling the translocation velocity via DNA charge inversion (CI). It is noteworthy that this peculiarity has been subsequently observed by translocation experiments~\cite{e21}. CI being induced by ion correlations, MF theories are unable to predict this effect.

The optimization of polymer translocation necessitates, in addition to a low translocation velocity,  the fast capture of anionic polymers by negatively charged silicon-based nanopores~\cite{e19}. Thus, the technical challenge consists in overcoming the repulsive electrostatic coupling between the polymer and the pore surface charges. At the theoretical level, this issue can be addressed only by a translocation model accounting for electrostatic polymer-pore interactions. Motivated by this point, we introduce herein the first translocation theory that includes both the direct electrostatic polymer-membrane coupling and ionic correlations absent in the Poisson-Boltzmann (PB) theory. Hence, the present formalism extends our purely electrohydrodynamic theory of Ref.~\cite{the16} to include the electrostatic interactions between the polymer and the membrane. The electrostatic part of our formalism is based on the one-loop (1l) theory of confined electrolytes~\cite{PodWKB,attard,netzcoun,jcp2,Buyuk2014}. We note that the accuracy of the 1l theory was previously verified by comparison with MC simulations of polyvalent ions confined to charged cylindrical pores~\cite{Buyuk2014}. In Ref.~\cite{the16}, the formalism was also shown to describe the experimentally measured ionic conductivity of nanopores with quantitative accuracy.

Our article is organized as follows. In Section~\ref{mod}, we extend our MF-level translocation theory of Ref.~\cite{mf} by incorporating the 1l-level drift transport theory of Ref.~\cite{the16} and the beyond-MF polymer-pore interaction potential derived in Ref.~\cite{correl}. The drift-driven transport regime of translocation events is considered in Section~\ref{drrev}. By direct comparisons with translocation experiments~\cite{e21}, we examine the electrohydrodynamic mechanism behind the correlation-induced DNA mobility reversal in solid-state pores. The electrohydrodynamics of polymer capture prior to the transport phase is investigated in Section~\ref{facil}. We show that in the typical case of strongly anionic solid-state pores in contact with a monovalent electrolyte bath, polymer capture is limited by the EO drag and the like-charge polymer-membrane repulsion. Added multivalent counterions remove the repulsive barrier, and activate the pore CI that reverses the direction of the EO flow to the trans side. However, the same multivalent ions also invert the DNA charge, turning the orientation of the electrophoretic (EP) drift to the cis side. We throughly characterize the resulting competition between the charge inverted EO and EP drag forces on DNA. We find that below (above) a characteristic polymer (membrane) charge strength, the inverted EO drag always dominates its EP counterpart and drives the polymer in the trans direction, therefore assisting the capture of the molecule by the pore. The facilitated polymer capture by polyvalent counterion addition is the key prediction of our work. The approximations of our model and possible improvements are discussed in Conclusions.

\begin{figure}
\includegraphics[width=.9\linewidth]{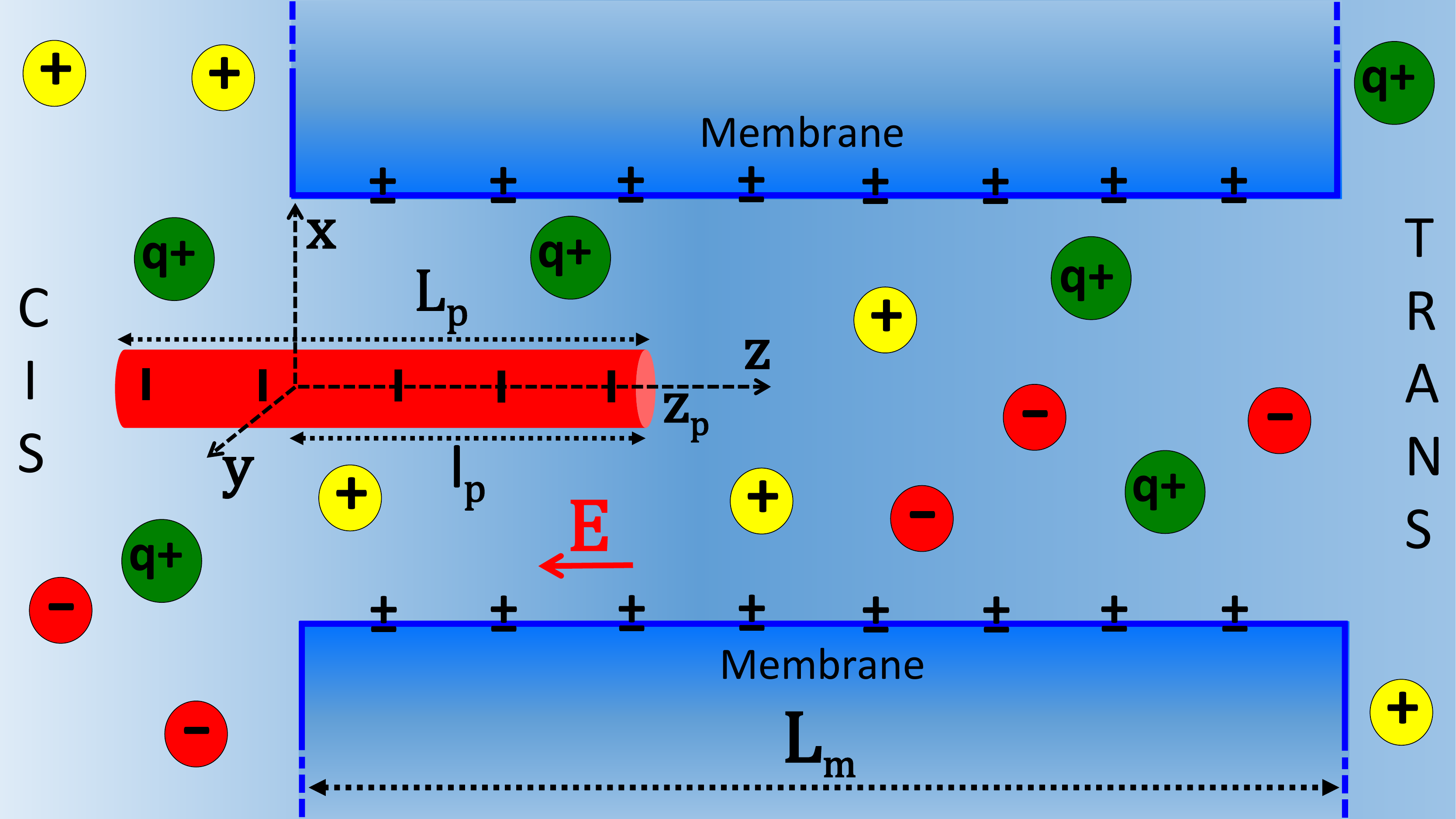}
\includegraphics[width=.9\linewidth]{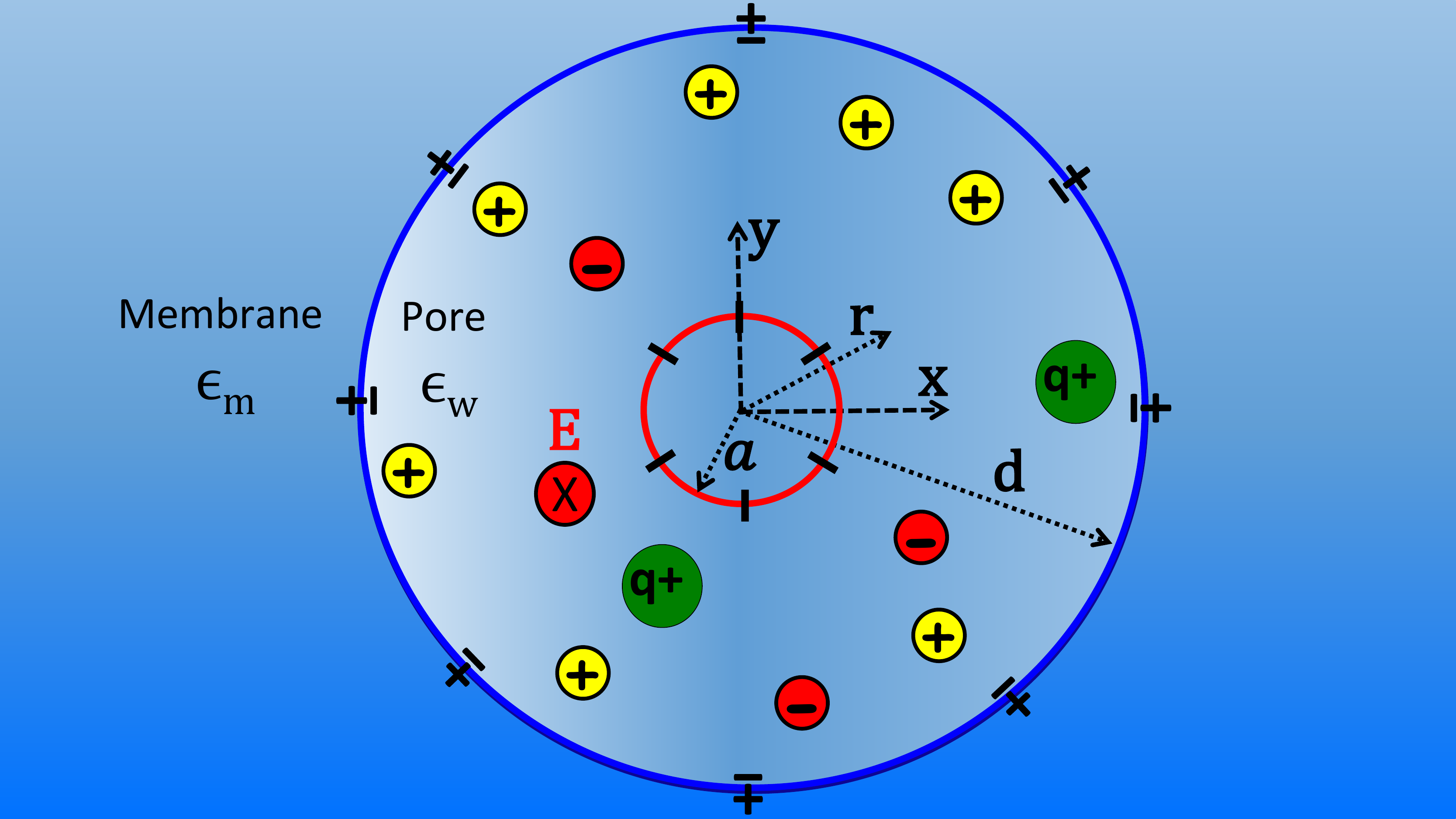}
\caption{(Color online) Schematic representation of the translocating polymer from the side (top plot) and the cross-section (bottom plot). The cylindrical polymer has length $L_p$, radius $a$, and negative fixed surface charge density $\sigma_p<0$. The pore is a cylinder with radius $d$, length $L_m$, and negative charge density $\sigma_m<0$. The membrane and pore dielectric permittivities are respectively $\e_m=2$ and $\e_w=80$. The polymer portion in the pore has length $l_p$ and its right end is located at $z=z_p$. Under the effect of the external electric field $\bE=-E\hat{u}_z$, translocation takes place along the $z$-axis.}
\label{fig1}
\end{figure}

\section{Model and Theory : Summary of previous results and inclusion of charge correlations}
\label{mod}

\subsection{Polymer translocation model}

The charge composition of the system is displayed in Fig.~\ref{fig1}. The nanopore is a cylinder of length $L_m$ and radius $d$, embedded in a membrane of dielectric permittivity $\e_m=2$. In our model, the discretely distributed fixed negative charges on the pore wall are taken into account by an effective continous charge distribution of surface density $\sigma_m<0$.  The nanopore is in contact with a bulk reservoir containing an electrolyte with dielectric permittivity $\e_w=80$ and temperature $T=300$ K. The electrolyte mixture is composed of $p$ ionic species. Each species $i$ has valency $q_i$ and reservoir concentration $\rho_{bi}$. 

The translocating polymer is a stiff cylinder with total length $L_p$ and radius $a$. The discrete charge distribution of the anionic polymer is approximated by a continous surface charge distrubution of density $\sigma_p<0$. For the sake of simplicity, we neglect off-axis polymer fluctuations and assume that the polymer and pore possess the same axis of symmetry. Hence, under the influence of the electric field $\bE=-E\hat{u}_z$ induced by the applied voltage $\Delta V=L_mE$, the translocation takes place along the $z$-axis. The polymer portion located inside the nanopore has length $l_p$. The position of its right end $z_p$ is the reaction coordinate of the translocation. In addition to the electric field, the translocating polymer is subject to the hydrodynamic drag force resulting from its interaction with the charged liquid, and the potential $V_p(z_p)$ induced by direct electrostatic polymer-pore interactions. 

\subsection{Electrohydrodynamic theory of polymer capture and transport}
\label{elth}

In this part, we review briefly the electrohydrodynamically augmented Smoluchowski formalism of Ref.~\cite{mf} and explain its extension beyond MF PB level. This polymer transport formalism is based on the Smoluchowski equation satisfied by the polymer probability density $c(z_p,t)$,
\be
\label{a1}
\frac{\partial c(z_p,t)}{\partial t}=-\frac{\partial J(z_p,t)}{\partial z_p},
\ee
with the polymer probability current 
\be
\label{a2}
J(z_p,t)=-D\frac{\partial c(z_p,t)}{\partial z_p}+c(z_p,t)v_p(z_p).
\ee
In Eq.~(\ref{a2}), the first term corresponds to Fick's law associated with the diffusive flux component. The diffusion coefficient $D$ for a cylindrical rigid polymer is
\be
\label{a3}
D=\frac{\ln(L_p/2a)}{3\pi\eta L_p\beta},
\ee
where we introduced the viscosity coefficient of water $\eta=8.91\times 10^{-4}\;\mathrm{Pa}\;\mathrm{s}$ and the inverse thermal energy $\beta=1/(k_BT)$~\cite{cyl1}. Then, the second term  of Eq.~(\ref{a2}) corresponds to the convective flux associated with the polymer motion at the velocity $v_p(z_p)$. In order to derive this velocity, we couple the Stokes and Poisson Eqs. 
\bea\label{st}
&&\eta\nabla^2_r u_c(r)-eE\rho_c(r)=0,\\
\label{pois}
&&\nabla^2_r \phi(r)+4\pi\ell_B\rho_c(r)=0,
\eea 
where $u_c(r)$ is the liquid velocity,  $\rho_c(r)$ the charge density, and $\phi(r)$ the electrostatic potential. Combining Eqs.~(\ref{st}) and~(\ref{pois}) and introducing the polymer mobility coefficient $\mu_e=\e_wk_BT/(e\eta)$, one obtains the relation $\nabla_r^2\left[u_c(r)+\mu_eE\phi(r)\right]=0$. Next, we integrate the latter equality and impose the no-slip condition at the pore wall $u_c(d)=0$ and the polymer surface $u_c(a)=v_p(z_p)$~\cite{rem1}. Finally, we use the force-balance relation on the polymer, 
\be\label{fb}
F_e+F_d+F_b=0, 
\ee
with the electric force $F_e=2\pi aL_p\sigma_p eE$, the hydrodynamic drag force $F_d=2\pi aL_p\eta u'_c(a)$, and the barrier-induced force $F_b=-V'_p(z_p)$. After some algebra, the solvent and polymer velocities follow as
\bea\label{a5}
u_c(r)&=&-\mu_eE\left[\phi(r)-\phi(d)\right]-\beta D_p(r)\frac{\partial V_p(z_p)}{\partial z_p},\\
\label{a6}
v_p(z_p)&=&v_{dr}-\beta D_p(a)\frac{\partial V_p(z_p)}{\partial z_p},
\eea
where we introduced the local diffusion coefficient
\be
\label{a7}
D_p(r)=\frac{\ln(d/r)}{2\pi\eta L_p\beta}.
\ee

The first component of Eq.~(\ref{a6}) is the drift velocity induced by the external field $\bE$,
\be
\label{a8}
v_{dr}=-\mu_e\left[\phi(a)-\phi(d)\right]E.
\ee
In Eq.~(\ref{a8}), the first term corresponds to the EP DNA velocity induced by the coupling between the electric field $\bE$ and the DNA molecule surrounded by its ionic cloud. The second term originates from the \textit{electroosmotic} (EO) flow composed of the ions attracted by the membrane charges. Finally, the second component of Eq.~(\ref{a6}) accounts for the alteration of the drift velocity~(\ref{a8}) by the interaction potential $V_p(z_p)$. 

The translocation rate will be calculated in the steady regime of Eq.~(\ref{a1}) where the probability current~(\ref{a2}) is constant, i.e. $J(z_p,t)=J_0$. First, we introduce the effective polymer potential
\be
\label{a9}
U_p(z_p)=\frac{D_p(a)}{D}V_p(z_p)-\frac{v_{dr}}{\beta D}z_p
\ee
and substitute the velocity~(\ref{a6}) into Eq.~(\ref{a2}). Then,  we integrate Eq.~(\ref{a2}) by imposing the polymer density at the cis side $c(z_p=0)=c_{out}$ and the absorbing boundary condition at the trans side  $c(z_p=L_m+L_p)=0$~\cite{rem2}. The translocation rate defined as the ratio of the polymer current and density at the pore entrance follows as
\be\label{a11}
R_c=\frac{D}{\int_0^{L_m+L_p}\mathrm{d}z\;e^{\beta U_p(z)}}.
\ee
The rate $R_c$ corresponds to the characteristic speed at which a successfull translocation takes place. The form of the potential~(\ref{a9}) indicates that the polymer conductivity of the pore is determined by the competition between the voltage-induced drift and electrostatic polymer-pore interactions. In the \textit{drift regime} characterized by negligible interactions, i.e. $V_p(z_p)\ll k_BT$, Eq.~(\ref{a11}) becomes
\be\label{34}
R_c\approx\frac{v_{dr}}{1-e^{-v_{dr}(L_m+L_p)/D}}\approx v_{dr},
\ee
where the second equality holds for high electric fields and a positive drift velocity.

The translocation rate~(\ref{a11}) depends on the effective potential $U_p(z_p)$ introduced in Eq.~(\ref{a9}). In Section~\ref{elpot}, the polymer-pore interaction potential $V_p(z_p)$ appearing in the first term of Eq.~(\ref{a9}) will be derived in terms of the polymer grand potential previously computed in Ref.~\cite{correl}. The second component of Eq.~(\ref{a9}) includes the drift velocity~(\ref{a8}) depending on the pore potential $\phi(r)$. In the present work, the potential $\phi(r)$ will be evaluated within the 1l theory of electrostatic interactions that improves the PB theory by including charge correlations~\cite{PodWKB,attard,netzcoun,jcp2,Buyuk2014}. According to the 1l theory, the pore potential $\phi(r)$ is composed of two contributions, $\phi(r)=\phi_0(r)+\phi_c(r)$. The MF component $\phi_0(r)$ solves the PB equation $\nabla^2_r \phi_0(r)+4\pi\ell_B\sum_{i=1}^pq_i\rho_{bi}e^{-q_i\phi_0(r)}=0$. The additional component $\phi_c(r)$ brings correlation corrections. The computation of these two potential components is explained in Appendix~\ref{avpot}. 

\subsection{Computing the beyond-MF polymer-pore interaction potential $V_p(z_p)$}
\label{elpot}

The interaction potential $V_p(z_p)$ will be computed by taking into account exclusively the interaction between the membrane and the polymer portion in the pore. Thus, the evaluation of the potential $V_p(z_p)$ requires the knowledge of the polymer grand potential $\Delta\Omega_p(l_p)$ corresponding to the electrostatic cost of polymer penetration by the length $l_p$ into the pore. First, we summarize the derivation of this grand potential within the 1l-test charge theory~\cite{correl}. The polymer charge structure will be approximated by a charged line with density $\tau=2\pi a\sigma_p$. The grand potential is composed of two components,
\be
\label{1}
\Delta\Omega_p(l_p)=\Omega_{mf}(l_p)+\Delta\Omega_s(l_p).
\ee

The first term of Eq.~(\ref{1}) corresponds to the MF grand potential associated with the direct polymer-pore charge coupling, $\beta\Omega_{mf}(l_p)=\int\mathrm{d}\br\sigma_p(\br)\phi_m(r)$. The integral includes the charge density function of the linear polymer $\sigma_p(\br)=\tau\theta(z)\theta(l_p-z)\delta(r)/(2\pi r)$ where $\theta(x)$ and $\delta(x)$ are respectively the Heaviside step and Dirac delta functions~\cite{math}, and the electrostatic potential $\phi_m(r)$ induced exclusively by the membrane charges. The potential $\phi_m(r)$ solves the PB Eq. 
\be\label{2II}
\frac{1}{4\pi\ell_Br}\partial_r\left[r\partial_r\phi_m(r)\right]+\sum_{i=1}^p\rho_{bi}q_ie^{-q_i\phi_m(r)}=-\sigma_m\delta(r-d),
\ee
where $\ell_B\approx7$ {\AA} is the Bjerrum length. In Ref.~\cite{correl}, the solution of Eq.~(\ref{2II}) was derived within a Donnan potential approximation in the form
\be
\label{2III}
\phi_m(r)=\phi_d+\frac{4\pi\ell_B\sigma_m}{\kappa_d}
\left[\frac{\mathrm{I}_0(\kappa_dr)}{\mathrm{I}_1(\kappa_dd)}-\frac{2}{\kappa_dd}\right],
\ee
where $\phi_d$ is the constant Donnan potential solving the equation $\sum_{i=1}^p\rho_{bi}q_ie^{-q_i\phi_d}=-2\sigma_m/d$ and the Donnan screening parameter reads $\kappa_d^2=4\pi\ell_B\sum_{i=1}^p\rho_{bi}q_i^2e^{-q_i\phi_d}$. Finally, the MF grand potential follows as $\beta\Omega_{mf}(l_p)=l_p\psi_{mf}$, with the MF grand potential per length
\be\label{6II}
\psi_{mf}=\tau\phi_d+\tau\frac{4\pi\ell_B\sigma_m}{\kappa_d}\left[\frac{1}{\mathrm{I}_1(\kappa_dd)}-\frac{2}{\kappa_dd}\right].
\ee
For a negatively charged membrane, the MF grand potential $\Omega_{mf}(l_p)$ is positive and rises linearly with the penetration length $l_p$. This reflects the hindrance of the polymer capture by repulsive polymer-pore interactions. 

The second term of Eq.~(\ref{1}) is the polymer self-energy difference between the pore and the bulk reservoir,
\be
\label{7}
\beta\Delta\Omega_s(l_p)=\frac{1}{2}\int\mathrm{d}\br\mathrm{d}\br'\sigma_p(\br)\left[v(\br,\br')-v_b(\br-\br')\right]\sigma_p(\br').
\ee
Eq.~(\ref{7})  brings electrostatic correlations to the polymer grand potential~(\ref{1}). The electrostatic propagator in the pore $v(\br,\br')$ solves the kernel equation
\be\label{7II}
\left[\nabla\e(r)\nabla-\e(r)\kappa^2(r)\right]v(\br,\br')=-\frac{e^2}{k_BT}\delta(\br-\br'),
\ee
with the dielectric permittivity profile $\e(r)=\e_w\theta(d-r)+\e_m\theta(r-d)$ and the local screening parameter 
\be\label{kl}
\kappa^2(r)=4\pi\ell_B\sum_{i=1}^p\rho_{bi}q_i^2e^{-q_i\phi_m(r)}\theta(d-r). 
\ee
Eq.~(\ref{7}) also includes the bulk propagator corresponding to the screened Debye-H\"{u}ckel potential $v_b(\br)=\ell_Be^{-\kappa_b|\br|}/|\br|$, with the bulk screening parameter
\be\label{8}
\kappa_b^2=4\pi\ell_B\sum_{i=1}^p\rho_{bi}q_i^2.
\ee

In Ref.~\cite{correl}, Eq.~(\ref{7II}) was solved within a Wentzel-Kramers-Brillouin (WKB) scheme and the resulting self-energy~(\ref{7}) was obtained in the form $\beta\Delta\Omega_s(l_p)=l_p\psi_s(l_p)$, with the self-energy density
\bea\label{9II}
\psi_s(l_p)&=&\ell_B\tau^2\int_{-\infty}^\infty\mathrm{d}k\frac{2\sin^2(kl_p/2)}{\pi l_pk^2}\\
&&\hspace{1.9cm}\times\left\{-\ln\left[\frac{p(0)}{p_b}\right]+\frac{Q(k)}{P(k)}\right\}.\nonumber
\eea
Eq.~(\ref{9II}) includes the auxiliary functions
\bea\label{10}
Q(k)&=&2p^3(d)dB_0(d)\mathrm{K}_0\left(|k|d\right)\mathrm{K}_1\left[B_0(d)\right]\\
&&-2\gamma |k|dp^2(d)B_0(d)\mathrm{K}_1\left(|k|d\right)\mathrm{K}_0\left[B_0(d)\right]\nonumber\\
&&-\left[p^3(d)d-p^2(d)B_0(d)-\kappa(d)\kappa'(d)dB_0(d)\right]\nonumber\\
&&\hspace{3mm}\times\mathrm{K}_0\left(|k|d\right)\mathrm{K}_0\left[B_0(d)\right],\nonumber\\
\label{11}
P(k)&=&2p^3(d)dB_0(d)\mathrm{K}_0\left(|k|d\right)\mathrm{I}_1\left[B_0(d)\right]\\
&&+2\gamma |k|dp^2(d)B_0(d)\mathrm{K}_1\left(|k|d\right)\mathrm{I}_0\left[B_0(d)\right]\nonumber\\
&&+\left[p^3(d)d-p^2(d)B_0(d)-\kappa(d)\kappa'(d)dB_0(d)\right]\nonumber\\
&&\hspace{3mm}\times\mathrm{K}_0\left(|k|d\right)\mathrm{I}_0\left[B_0(d)\right],\nonumber
\eea
with the modified Bessel functions $\mathrm{I}_n(x)$ and $\mathrm{K}_n(x)$\cite{math}, the parameters $\gamma=\e_m/\e_w$ and $p_b=\sqrt{\kappa_b^2+k^2}$, and
\be
\label{12II}
p(r)=\sqrt{\kappa^2(r)+k^2}\hspace{2mm};\hspace{5mm}B_0(r)=\int_0^r\mathrm{d}r'p(r').
\ee
In Eq.~(\ref{9II}), the negative term accounts for the counterion excess induced by the fixed pore charges. This excess results in a more efficient screening of the polymer charges in the pore, which lowers the polymer grand potential and favours the polymer capture. At strong polymer charges, this negative term gives rise to the like-charge polymer pore attraction effect~\cite{correl}. Then, the positive term of Eq.~(\ref{9II}) embodies polymer-image charge interactions that increase  the polymer grand potential and act as a barrier limiting the polymer penetration.

The electrostatic potential landscape $V_p(z_p)$ is related to the polymer grand potential~(\ref{1}) by the equality $V_p(z_p)=\Delta\Omega_p\left[l_p(z_p)\right]$.  Defining the auxiliary lengths
\be
\label{lmp}
L_-=\mathrm{min}(L_m,L_p)\;;\hspace{5mm} L_+=\mathrm{max}(L_m,L_p),
\ee
the polymer penetration length $l_p$ can be expressed in terms of the polymer position $z_p$ as
\bea
\label{lpzp}
l_p(z_p)&=&z_p\theta(L_--z_p)+L_-\theta(z_p-L_-)\theta(L_+-z_p)\nonumber\\
&&+(L_p+L_m-z_p)\theta(z_p-L_+).
\eea
Thus, the potential profile $V_p(z_p)$ finally becomes
\bea\label{vp}
V_p(z_p)&=&\Delta\Omega_p(l_p=z_p)\theta(L_--z_p)\\
&&+\Delta\Omega_p(l_p=L_-)\theta(z_p-L_-)\theta(L_+-z_p)\nonumber\\
&&+\Delta\Omega_p(l_p=L_p+L_m-z_p)\theta(z_p-L_+).\nonumber
\eea
The first term of Eq.~(\ref{vp}) corresponds to the energetic cost for polymer penetration during the capture regime $z_p\leq L_-$. The second term coincides with the transport regime $L_-\leq z_p\leq L_+$ where the fully penetrated polymer diffuses at the drift velocity $v_p(z_p)=v_{dr}$. Finally, the third term corresponds to the exit phase at $z_p\geq L_+$.

\section{Results and discussion}
\label{res}

Artificially fabricated solid-state pores with chosen characteristics and improved solidity offer an efficient approach to biopolymer sensing~\cite{Tapsarev}. These silicon-based membrane pores of large radius $d\gg1$ nm carry negative fixed charges~\cite{e19} and therefore strongly adsorb the multivalent cations added to the reservoir. We examine here the polymer capture and transport properties of such pores filled with multivalent cations that drive the system beyond the MF electrohydrodynamic regime. 

\begin{figure}
\includegraphics[width=.9\linewidth]{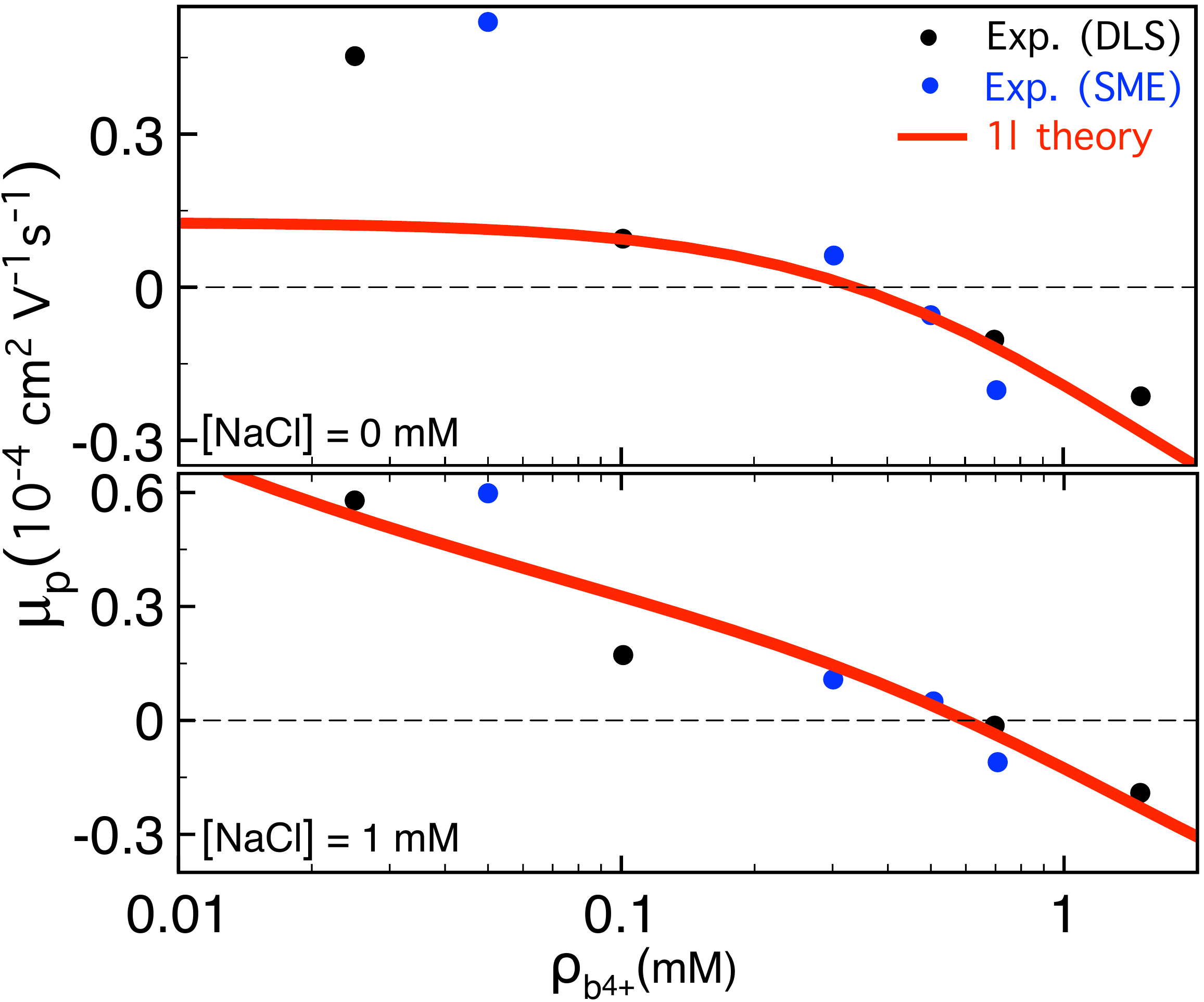}
\caption{(Color online) Polymer mobility $\mu_p=v_{dr}/E$ versus $\mbox{Spm}^{4+}$ density $\rho_{b4+}$ in the NaCl$+\mbox{SpmCl}_4$ mixture with the monovalent cation density $\rho_{b+}=0$ mM (top) and $1$ mM (bottom). Solid curves:  1l result from Eq.~(\ref{a8}). Dots: experimental data of Ref.~\cite{e21} obtained from dynamic light scattering (DLS) and single molecule electrophoresis (SME)~\cite{rem3}. The polymer is a ds-DNA molecule with radius $a=1$ nm and effective charge density $\sigma_p=-0.12$ $e/\mbox{nm}^2$. The pore has radius $d=10$ nm and fixed charge density $\sigma_m=-0.006$ $e/\mbox{nm}^2$.}
\label{fig2}
\end{figure}

\subsection{DNA mobility reversal : theory versus experiments}
\label{drrev}

We reconsider here the effect of DNA velocity reversal~\cite{the16} and present our  first comparison with translocation experiments~\cite{e21}. Fig.~\ref{fig2} displays the DNA mobility $\mu_p=v_{dr}/E$ versus the spermine ($\mbox{Spm}^{4+}$) density in the NaCl$+\mbox{SpmCl}_4$ solution at two monovalent cation density values. The result corresponds to the transport regime of the translocation process where the captured polymer diffuses at the drift velocity $v_p(z_p)=v_{dr}$. The theoretical prediction of Eq.~(\ref{a8}) is displayed together with the experimental data of Ref.~\cite{e21} obtained from dynamic light scattering (DLS) and single molecule electrophoresis (SME)~\cite{rem3}. The effective membrane and polymer charge densities were adjusted in order to obtain the best fit with the magnitude of the experimental data (see the caption) while the pore radius was set to the value $d=10$ nm located in the characteristic range of solid-state pores.

\begin{figure}
\includegraphics[width=1\linewidth]{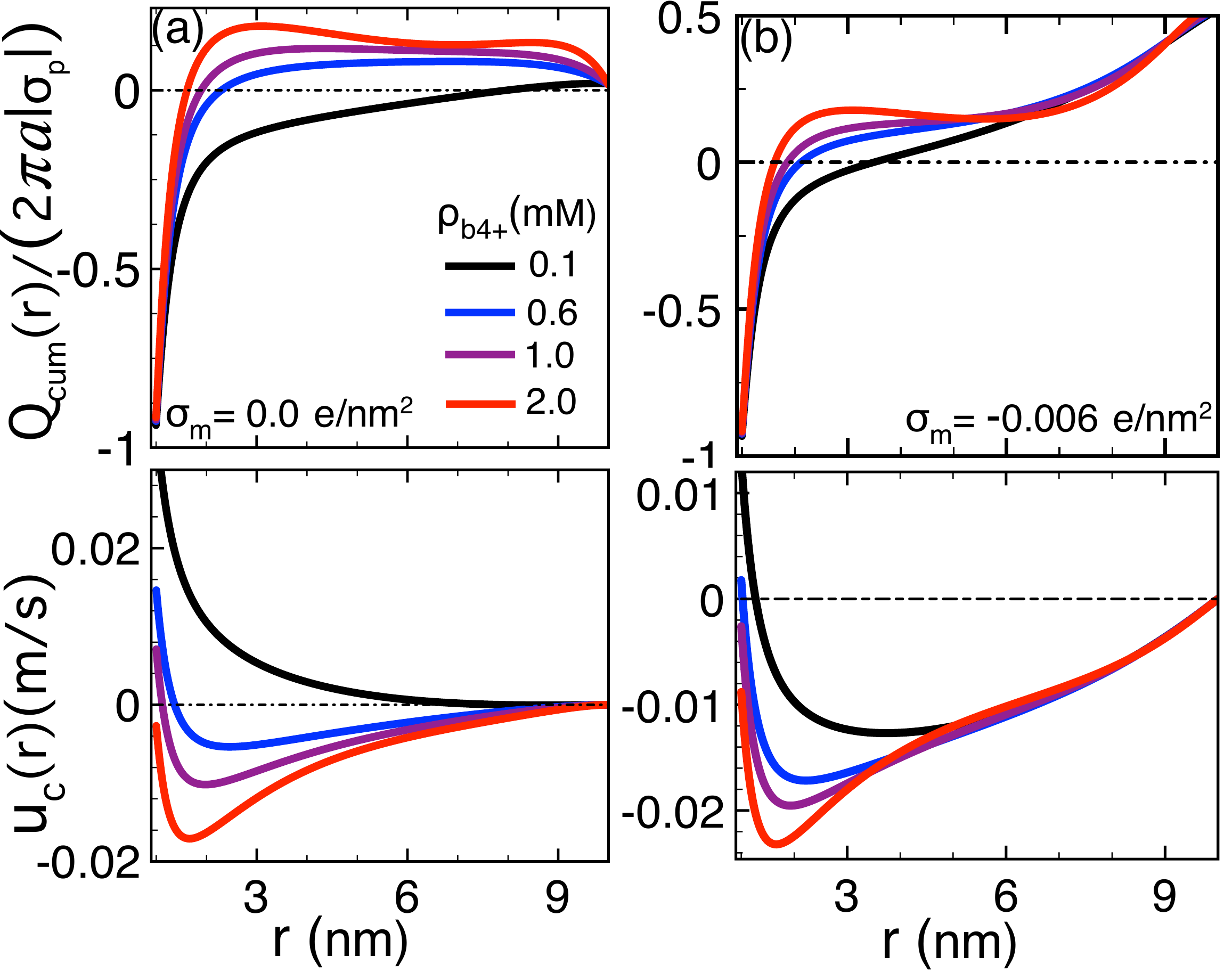}
\caption{(Color online) Adimensional cumulative charge density $Q_{cum}(r)/(2\pi a|\sigma_p|)$ (top plots) and convective liquid velocity profile $u_c(r)$ (bottom plots) in (a) neutral pores and (b) weakly charged pores with density $\sigma_m=-0.006$ $e/\mbox{nm}^2$. The external voltage is $\Delta V=120$ mV, the pore length $L_m=34$ nm, and the monovalent cation density $\rho_{b+}=1$ mM. The other parameters are the same as in Fig.~\ref{fig2}.}
\label{fig3}
\end{figure}

Fig.~\ref{fig2} shows that the low $\mbox{Spm}^{4+}$ density regime is characterized by a positive DNA mobility indicating the motion of the anionic polymer oppositely to the field $\bE$. Upon the increment of the $\mbox{Spm}^{4+}$ density, the mobility drops ($\rho_{b4+}\uparrow\mu_p\downarrow$) and turns to negative, i.e. DNA changes its direction and translocates parallel with the field $\bE$. Then, the comparison of the top and bottom plots shows that monovalent salt increases both the mobility ($\rho_{b+}\uparrow\mu_p\uparrow$) and the critical $\mbox{Spm}^{4+}$ density $\rho_{b4+}^*$ for mobility reversal ($\rho_{b+}\uparrow\rho^*_{b4+}\uparrow$). Within the experimental uncertainty, our theory can account for these characteristics with reasonable accuracy, except at low $\mbox{Spm}^{4+}$ densities where the mobility data is underestimated.

\begin{figure*}
\includegraphics[width=.9\linewidth]{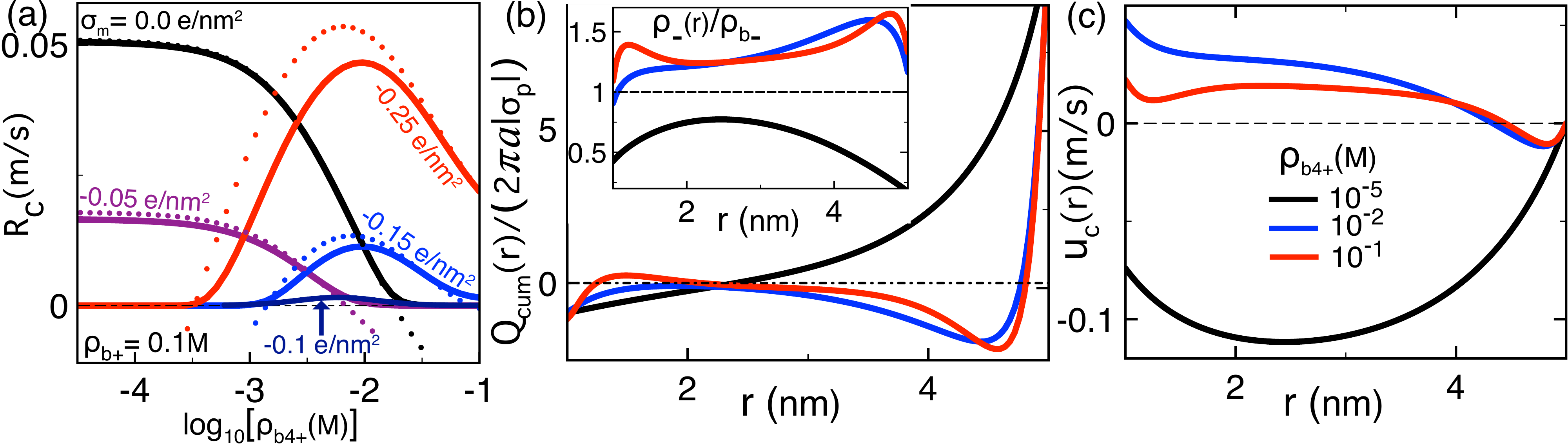}
\caption{(Color online) (a) Translocation rate $R_c$ (solid curves) and drift velocity $v_{dr}$ (dots) against the $\mbox{Spm}^{4+}$ density at various membrane charge values. (b) Adimensional cumulative charge (main plot) and $\mbox{Cl}^-$ density (inset), and (c) liquid velocity $u_c(r)$ at the membrane charge $\sigma_m=-0.25$ $e/\mbox{nm}^2$  and various $\mbox{Spm}^{4+}$ densities given in the legend. The polymer length is $L_p=10$ nm, the pore radius $d=5$ nm, and the $\mbox{Na}^+$ density $\rho_{b+}=0.1$ M. The other parameters are the same as in Fig.~\ref{fig3}.}
\label{fig4}             
\end{figure*}

In order to illustrate the mechanism behind the mobility reversal, in Fig.~\ref{fig3}, we plotted the cumulative charge  
\be\label{cum}
Q_{cum}(r)=2\pi\int_a^r\mathrm{d}r'r'\left[\rho_c(r')+\sigma_p(r')\right]
\ee
corresponding to the net charge of the DNA-counterion cloud complex (top plots), with the local charge density $\rho_c(r)$ given in Appendix~\ref{avpot}. We also reported the liquid velocity $u_c(r)$ of Eq.~(\ref{a5}) in the translocation regime $L_-<z_p<L_+$ where the barrier component vanishes (bottom plots). To consider first the effect of electrophoresis only, in Fig.~\ref{fig3}(a), we turned off the EO flow by setting $\sigma_m=0$. At the lowest $\mbox{Spm}^{4+}$ density $\rho_{b4+}=0.1$ mM (black curve), the MF-level counterion binding to DNA results in a negative liquid charge $Q_{cum}(r)\leq0$. As a result, the DNA and its counterion cloud move oppositely to the external field $\bE$, i.e. $u_c(r)\geq0$ and $v_{dr}=u_c(a)>0$. 

At the larger  $\mbox{Spm}^{4+}$ densities $\rho_{b4+}=0.6$ mM (blue curves) and $1.0$ mM (purple curves) with enhanced charge correlations, far away from the DNA surface, the cumulative charge density switches from negative to positive. This is the signature of DNA CI. As a result, in the same region, the solvent changes its direction and moves parallel with the field $\bE$, i.e. $u_c(r)<0$. However, at the corresponding $\mbox{Spm}^{4+}$  densities where CI is not strong enough, the drag force on DNA is not sufficient to compensate the coupling between the electric field and the DNA charges. Consequently, the DNA and the liquid in its close vicinity continue to move oppositely to the field $\bE$, i.e. $v_{dr}>0$. The further increase of the $\mbox{Spm}^{4+}$ density to $\rho_{b4+}=2.0$ mM (red curves) amplifies the inverted liquid charge. This results in an enhanced hydrodynamic drag force that dominates the electric force on DNA and reverses the mobility of the molecule, i.e. $v_{dr}<0$.

Hence, a strong enough CI can solely invert the DNA mobility. To obtain analytical insight into this causality, we integrate the Stokes Eq.~(\ref{st}) to get
\be\label{int}
u'_c(r)=\frac{eQ_{cum}(r)E}{2\pi r\eta}.
\ee
Eq.~(\ref{int}) is a macroscopic force-balance relation equating the drag force $F_{hyd}=2\pi rL\eta u'_c(r)$ and the electric force $F_{el}=eQ_{cum}(r)LE$ on the polymer-liquid complex located within the arbitrary cylindrical surface $S=2\pi rL$. In agreement with Fig.~\ref{fig3}, Eq.~(\ref{int}) states that charge reversal gives rise to the minimum of the liquid velocity $u_c(r)$. This minimum should be however deep enough for the drift velocity $v_{dr}=u_c(a)$ to become negative. This explains the necessity to have a strong enough CI for the occurrence of the DNA mobility reversal. 

The additional effect of the EO flow is displayed in Fig.~\ref{fig3}(b) including the finite membrane charge of Fig.~\ref{fig2}.  The comparison of Figs.~\ref{fig3}(a) and (b) shows that the cations attracted by the membrane charges enhance the positive cumulative charge density. The resulting EO flow lowers the liquid velocity and reduces the critical $\mbox{Spm}^{4+}$  density for mobility inversion, i.e. $|\sigma_m|\uparrow\rho_{b4+}^*\downarrow$. Hence, the DNA velocity reversal in Fig.~\ref{fig2} is mainly due to CI whose effect is augmented by the EO flow.

\subsection{Facilitated polymer capture by inverted EO flow}
\label{facil}
\subsubsection{Effect of ion concentration and pore surface charge}  
\label{mulcap}

Having scrutinized the effect of $\mbox{Spm}^{4+}$ molecules on the DNA drift velocity,  we characterize the role played by correlations on polymer capture. Fig.~\ref{fig4}(a) displays the alteration of the translocation rate $R_c$ (solid curves) and drift velocity $v_{dr}$ (dots) by $\mbox{Spm}^{4+}$ molecules. Due to the high $\mbox{Na}^+$ density $\rho_{b+}=0.1$ M, DNA-pore interactions are strongly screened, i.e. $V_p(z_p)\ll k_BT$. Thus, the system is located in the drift-driven regime of Eq.~(\ref{34}) where $R_c$ closely follows the drift velocity $v_{dr}$.  

In weakly anionic pores $|\sigma_m|\lesssim0.1$ $e/\mbox{nm}^2$ (black and purple curves),  the addition of $\mbox{Spm}^{4+}$ molecules monotonically lowers $R_c$ and turns the drift velocity $v_{dr}$ from positive to negative. Thus,  $\mbox{Spm}^{4+}$ molecules hinder polymer capture via the EP mobility reversal induced by  DNA CI. Then, one notes that for $|\sigma_m|\lesssim0.1$ $e/\mbox{nm}^2$, $R_c$ is also reduced by the increase of the membrane charge, i.e. $|\sigma_m|\uparrow R_c\downarrow$. This stems from the onset of the EO flow opposing the EP motion of DNA~\cite{the3,aks4, mf}.

In the stronger membrane charge regime $|\sigma_m|\gtrsim0.1$ $e/\mbox{nm}^2$, this situation is reversed; polymer capture is significantly facilitated by the addition of $\mbox{Spm}^{4+}$ molecules ($\rho_{b4+}\uparrow R_c\uparrow$) up to the density $\rho_{b4+}\sim0.01$ M where $R_c$ reaches a peak and decays beyond this value ($\rho_{b4+}\uparrow R_c\downarrow$). In addition, translocation rates rise with the membrane charge strength, i.e. $|\sigma_m|\uparrow R_c\uparrow$. Thus, in the presence of a sufficient amount of $\mbox{Spm}^{4+}$ molecules, fixed negative pore charges of high density promote the capture of the like-charged polymer. 
\begin{figure}
\includegraphics[width=1.0\linewidth]{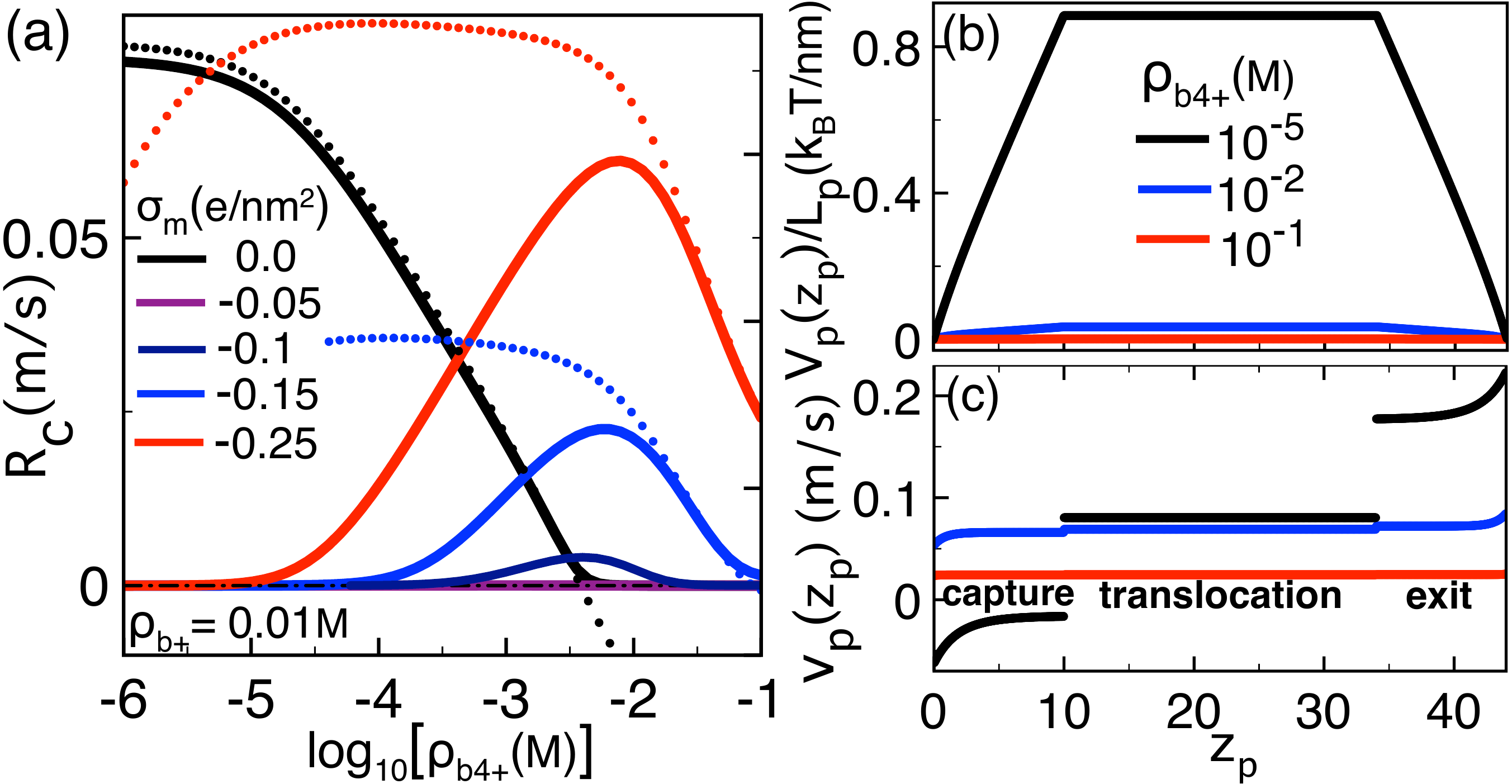}
\caption{(Color online) (a) Translocation rate $R_c$ (solid curves) and drift velocity $v_{dr}$ (dots) against the $\mbox{Spm}^{4+}$ concentration at various membrane charge values. (b) Polymer-pore interaction potential $V_p(z_p)$ and (c) velocity profile $v_p(z_p)$ at the membrane charge $\sigma_m=-0.25$ $e/\mbox{nm}^2$ and various $\mbox{Spm}^{4+}$ concentration values. $\mbox{Na}^+$ concentration is $\rho_{b+}=0.01$ M. The other parameters are the same as in Fig.~\ref{fig4}.}
\label{fig5}
\end{figure}

The enhancement of the translocation rates by polyvalent cations originates from the inversion of the EO flow. Fig.~\ref{fig4}(b) indicates that the increase of the $\mbox{Spm}^{4+}$ density from $\rho_{b4+}\approx10^{-5}$ M to $10^{-2}$ M results in the CI of the pore wall as the latter attracts like-charged $\mbox{Cl}^-$ ions (inset) and the cumulative charge density switches from positive to negative (main plot). Fig.~\ref{fig4}(c) shows that due to hydrodynamic drag, this negatively charged EO flow moving oppositely to the field $\bE$ turns the DNA velocity $v_{dr}=u_c(a)$ from negative to positive and assists the capture of the molecule by the pore. The anionic pore charge and streaming current reversal by polyvalent cations has been previously observed in nanofluidic experiments~\cite{Heyden2006}. 

In Fig.~\ref{fig4}(b), one sees that the further increase of the $\mbox{Spm}^{4+}$ density from $\rho_{b4+}=10^{-2}$ M  to $10^{-1}$ M weakens the $\mbox{Cl}^-$ attraction and the inverted cumulative charge density close to the pore wall. Fig.~\ref{fig4}(c) shows that this lowers the inverted EO flow velocity,  leading to the decay of the drift velocities and translocation rates in Fig.~\ref{fig4}(a) ($\rho_{b4+}\uparrow v_{dr}\downarrow R_c\downarrow$). The dissipation of the pore CI stems from the screening of the pore potential $\phi(r)$ by $\mbox{Spm}^{4+}$ molecules of large concentration. This effect has been equally observed in the experiments of Ref.~\cite{Heyden2006}. 

We consider now the opposite regime of dilute monovalent salt and set $\rho_{b+}=0.01$ M. Fig.~\ref{fig5}(a) shows that at the pore charges $|\sigma_m|\gtrsim0.1$ $e/\mbox{nm}^2$ and low $\mbox{Spm}^{4+}$ densities, the charge inverted EO flow results in a positive drift velocity $v_{dr}>0$ but the translocation rate is vanishingly small, i.e. $R_c\ll v_{dr}$. The loss of correlation between $R_c$ and  $v_{dr}$  originates from electrostatic DNA-pore interactions that become relevant at dilute salt and drive the system to the barrier-driven regime. Indeed, Fig.~\ref{fig5}(b) shows that at the $\mbox{Spm}^{4+}$ density $\rho_{b4+}=10^{-5}$ M, the like-charge polymer-pore repulsion results in a significant barrier $V_p(z_p)/L_p\sim$ $k_BT/$nm. In Fig.~\ref{fig5}(c), one sees that this barrier leads to a negative velocity $v_p(z_p)<0$ at the pore entrance $z_p<10$ nm, thus hindering the polymer capture. Due to the negative term of the self-energy~(\ref{9II}), the increment of the $\mbox{Spm}^{4+}$ density from $\rho_{b4+}=10^{-5}$ M  to $10^{-2}$ M enhances the screening ability of the pore and removes the electrostatic barrier $V_p(z_p)$. As a result, the capture velocity turns to positive and results in the rise of the  translocation rates  ($\rho_{b4+}\uparrow R_c\uparrow$) in Fig.~\ref{fig5}(a). 

\begin{figure}
\includegraphics[width=1.0\linewidth]{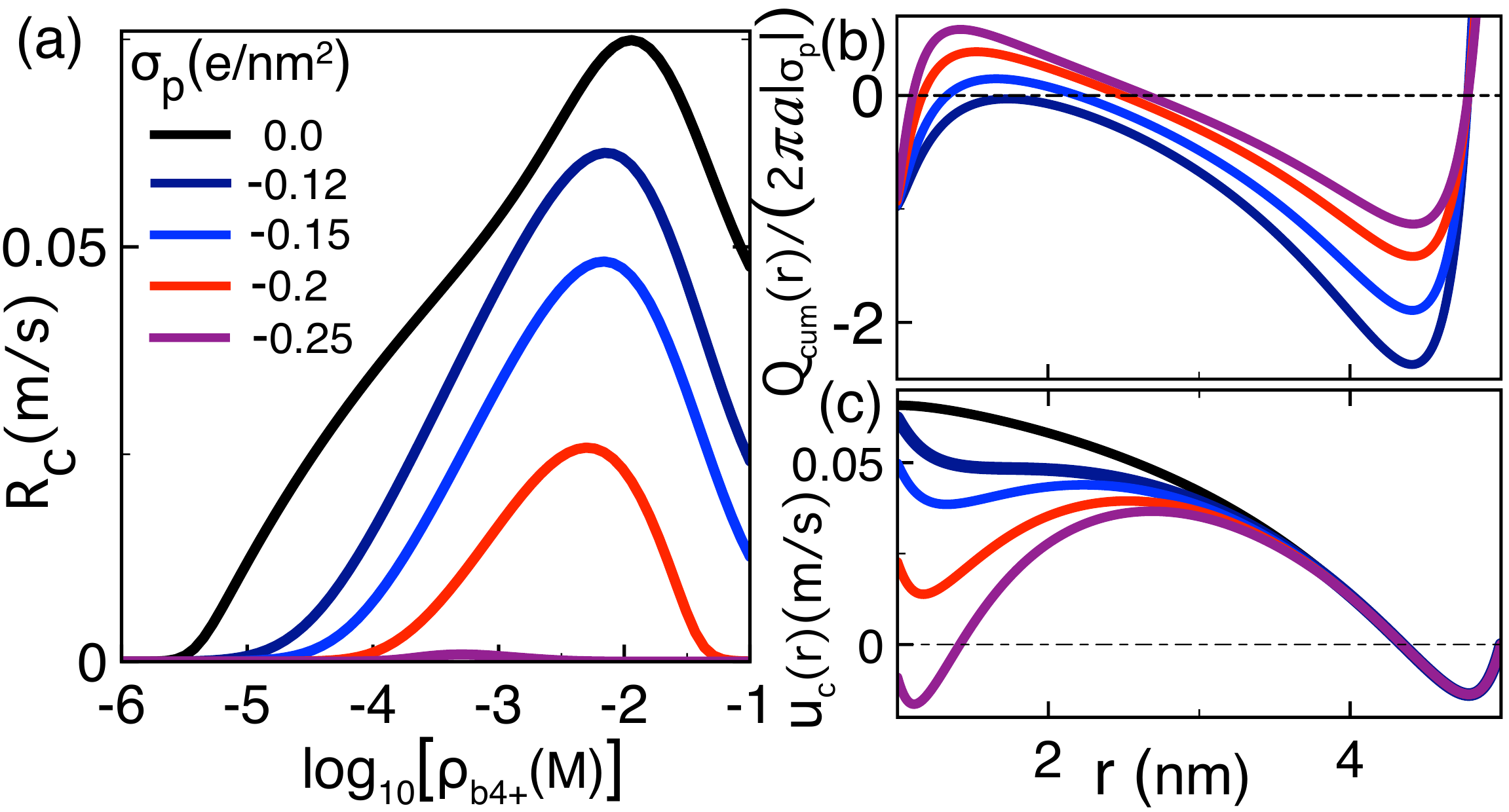}
\caption{(Color online) (a) Translocation rate against the $\mbox{Spm}^{4+}$ density. (b) Cumulative charge and (c) liquid velocity profile at the $\mbox{Spm}^{4+}$ density $\rho_{b4+}=10^{-2}$ M. The polymer charge density for each curve is indicated in (a). The membrane charge is $\sigma_m=-0.25$ $e/\mbox{nm}^2$ and the $\mbox{Na}^+$ density $\rho_{b+}=0.01$ M. The other parameters are the same as in Fig.~\ref{fig4}.}
\label{fig6}
\end{figure}

\subsubsection{Effect of polymer charge and sequence length}
\label{finsz}

We found that in strong salt conditions where polymer translocation is drift-driven, the enhancement of polymer capture by polyvalent cations is induced by the EO flow reversal. In dilute salt where the system is in the barrier-driven regime, facilitated polymer capture by $\mbox{Spm}^{4+}$ molecules originates from the removal of the electrostatic barrier. We investigate now the effect of the polymer charge strength on polymer capture. Interestingly, Fig.~\ref{fig6}(a) shows that in the presence of $\mbox{Spm}^{4+}$ molecules, polymer capture is hindered by the molecular charge, i.e. $|\sigma_p|\uparrow R_c\downarrow$. This peculiarity results from the polymer CI. Figs.~\ref{fig6}(b) and (c) indicate that the increase of the polymer charge strength amplifies the inverted cumulative charge ($|\sigma_p|\uparrow Q_{cum}(r)\uparrow$) and lowers the drift velocity $v_{dr}=u_c(a)$. Beyond the charge density $|\sigma_p|\approx0.2$ $e/\mbox{nm}^2$,  the reversed EP mobility takes over the inverted EO drag and turns the drift velocity to negative (purple curves). The resulting anticorrelation between the polymer charge and translocation rate ($|\sigma_p|\downarrow R_c\uparrow$) suggests that the inverted EO flow drag can be an efficient way to transport quasi-neutral polymers that cannot be controlled by electrophoresis.

\begin{figure}
\includegraphics[width=1.\linewidth]{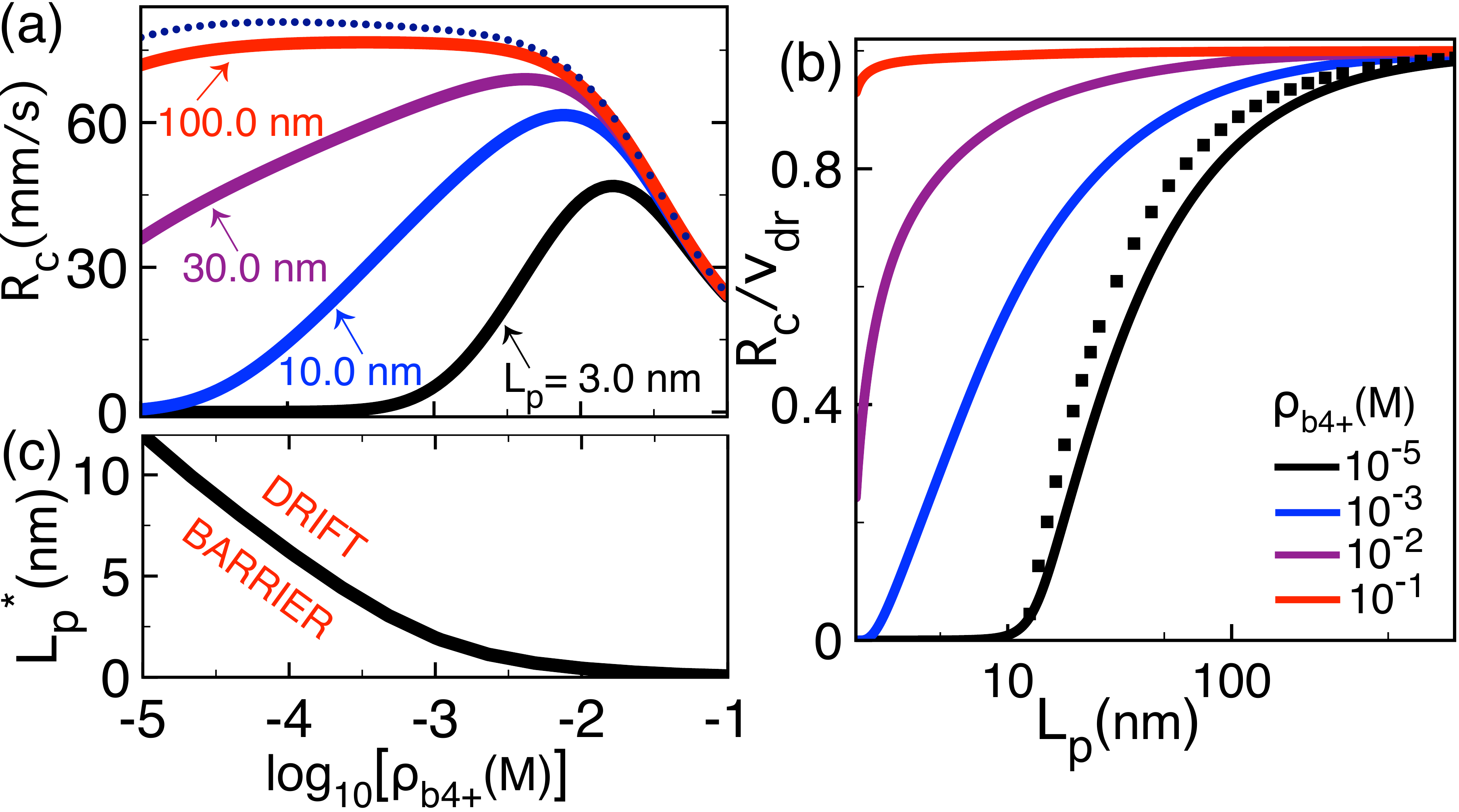}
\caption{(Color online) (a) Translocation rate $R_c$ at various polymer lengths (solid curves) and drift velocity $v_{dr}$ (dotted curve) against the $\mbox{Spm}^{4+}$ density. (b) Normalized translocation rate $R_c/v_{dr}$ against polymer length at various $\mbox{Spm}^{4+}$ densities. The squares at $\rho_{b4+}=10^{-5}$ M are from Eq.~(\ref{captr}).  (c) Critical polymer length~(\ref{lcr}) versus $\mbox{Spm}^{4+}$  density. Membrane charge is $\sigma_m=-0.25$ $e/\mbox{nm}^2$ and salt density $\rho_{b+}=0.01$ M. The other parameters are the same as in Fig.~\ref{fig4}.}
\label{fig10}
\end{figure}

Finally, we examine the effect of the molecular length on polymer capture. Fig.~\ref{fig10}(a) displays the alteration of the translocation rates $R_c$ by $\mbox{Spm}^{4+}$ molecules at various polymer lengths $L_p$. The increase of the length $L_p$ rises $R_c$ towards the drift velocity $v_{dr}$ and drives the system to the drift-driven regime. This peculiarity is also illustrated in Fig.~\ref{fig10}(b) displaying the translocation rate rescaled by the velocity $v_{dr}$; beyond a characteristic polymer length $L_p^*$, $R_c$ rises quickly ($L_p\uparrow R_c\uparrow$) and approaches the drift velocity ($R_c/v_{dr}\to1$) for $L_p\gg L_p^*$.

To explain this finite-size effect, we derive an analytical estimation of the translocation rate. Approximating the self-energy~(\ref{9II}) by its limit reached for a large polymer portion in the pore, $\kappa_{\rm b}l_{\rm p}\gg1$, one gets $\psi_s(l_p)\approx\psi_s$ with
\be
\label{selfth}
\psi_s=\ell_B\tau^2\left\{-\ln\left[\frac{\kappa(0)}{\kappa_{\rm b}}\right]+\frac{Q_0}{P_0}\right\},
\ee
where we introduced the geometric coefficients
\bea
\label{q0}
Q_0&=&2\kappa^2(d)dB(d)\mathrm{K}_1\left[B(d)\right]\\
&&-\left\{\kappa^2(d)d-\left[\kappa(d)+\kappa'(d)d\right]B(d)\right\}\mathrm{K}_0\left[B(d)\right],\nonumber\\
P_0&=&2\kappa^2(d)dB(d)\mathrm{I}_1\left[B(d)\right]\\
&&+\left\{\kappa^2(d)d-\left[\kappa(d)+\kappa'(d)d\right]B(d)\right\}\mathrm{I}_0\left[B(d)\right],\nonumber
\eea
and the function $B(r)=\int_0^r\mathrm{d}r'\kappa(r')$. Defining the characteristic lengths embodying the drift force $\lel=v_{dr}/D$ and barrier $\lb=D_p(a)\psi_t/D$, with the total barrier $\psi_t=\psi_{mf}+\psi_s$ and its MF component $\psi_{mf}$ given by Eq.~(\ref{6II}), the polymer potential~(\ref{a9}) becomes a piecewise linear function of the polymer position $z_p$, i.e. $\beta U_p(z_p)\approx\lb l_p(z_p)-\lel z_p$. To progress further, we approximate the translocation rate~(\ref{a11}) by the capture rate $R_c\approx D/\int_0^{L_-}\mathrm{d}z\;e^{\beta U_p(z)}$. Within this approximation whose accuracy will be shown below, one finds that in the barrier-dominated regime $\lambda_b>\lel$ corresponding to short sequences $L_p<L_p^*$, the capture rate increases exponentially with the polymer length,
\be
\label{bar}
R_c\approx v_{dr}\left(\frac{L_p^*}{L_p}-1\right)e^{-\lb L_-(1-L_p/L_p^*)},
\ee
with the characteristic sequence length 
\be\label{lcr}
L_p^*=\frac{\ln(d/a)\psi_t}{2\pi\eta\beta v_{dr}}
\ee
splitting the barrier and drift-driven regimes. Eq.~(\ref{bar}) is the Kramer's transition rate characterized by the barrier $\beta\Delta U=\lb L_-(1-L_p/L_p^*)$ to be overcome by the polymer in order to penetrate the pore. Then, in the drift-driven regime $\lel>\lb$ of long polymers $L_p>L_p^*$, $R_c$ rises and converges to the drift velocity $v_{dr}$ as an inverse linear function of the polymer length,
\be
\label{captr}
R_c\approx\left(1-\frac{L_p^*}{L_p}\right)v_{dr}.
\ee

Fig.~(\ref{fig10})(b) shows the reasonable accuracy of Eq.~(\ref{captr}) (see the black squares). The convergence to the drift regime with increasing length $L_p$ can be explained by the force-balance relation~(\ref{fb}); the electric force $F_e$ acts on the whole polymer with length $L_p$ while the barrier-induced force $F_b$ originates solely from the polymer portion in the pore. Thus, the rise of $L_p$ enhances the relative weight of the drift force with respect to the barrier. Then, in agreement with the $R_c-L_p$ curves, Fig.~\ref{fig10}(c) shows that the characteristic length $L_p^*$ drops with increasing $\mbox{Spm}^{4+}$ density, $\rho_{b4+}\uparrow L_p^*\downarrow$. This behavior is driven by the ionic solvation mechanism considered in Section~\ref{mulcap}; the addition of $\mbox{Spm}^{4+}$ molecules reduces the electrostatic barrier $\psi_t$ and shrinks the size of the barrier-driven region determined by Eq.~(\ref{lcr}). To summarize, facilitated polymer capture by inverted EO flow is achievable only with polymers longer than the characteristic length $L_p^*$. This length can be however reduced by $\mbox{Spm}^{4+}$ addition.

\section{Conclusions}

The predictive design of nanopore-based biosensing devices requires the complete characterization of polymer transport under experimentally realizable conditions. In this article, we introduced a beyond-MF translocation theory and characterized the polymer conductivity of nanopores in strong charge conditions where charge correlations lead to an unconventional polymer transport picture. Our main results are summarized below.

By comparison with translocation experiments, we investigated correlation effects on the electrophoretic mobility of DNA in solid-state pores. Fig.~\ref{fig2} shows that our theory can account for the mobility reversal by $\mbox{Spm}^{4+}$ molecules, as well as the rise of the DNA velocity and the characteristic $\mbox{Spm}^{4+}$ density for mobility inversion by monovalent salt. We also examined the causality between DNA CI and mobility reversal. Our results indicate the absence of one-to-one correspondence between these two phenomena. Indeed, the force-balance relation~(\ref{int}) shows that CI always leads to the reversal of the liquid velocity but this effect has to be strong enough to cause the reversal of the DNA mobility.

In the second part of our article, we considered the polymer capture regime prior to translocation. In the typical experimental configuration where a strongly anionic solid-state pore is in contact with a 1:1 electrolyte reservoir, polymer capture is limited by repulsive polymer-membrane interactions and the EO flow.  $\mbox{Spm}^{4+}$ molecules added to the reservoir suppress the repulsive interactions, and trigger the pore CI that reverses the direction of the EO flow. The inverted EO flow drags DNA towards the trans side and promotes its capture by the pore. We emphasize that an important challenge for serial biopolymer sequencing consists in enhancing the polymer capture speed from the reservoir. Thus, the facilitated polymer capture by $\mbox{Spm}^{4+}$ molecules is a key prediction of our work. Moreover, we found that due to the competition between the charge reversal of the EO and EP mobility components, the weaker the polymer charge, the more efficient the polymer capture driven by the inverted EO flow. Hence, this mechanism can be also useful for the transport of weakly charged polymers that cannot be controlled by electrophoresis. 

Due to the considerable complexity of the polymer translocation process, our theory involves approximations. In the solution of the electrostatic 1l and hydrodynamic Stokes equations, we neglected the finite length of the nanopore~\cite{Levin2006} as well as the discrete charge distribution on the DNA~\cite{Sung2013} and membrane surfaces. As these complications break the cylindrical symmetry of the model, their consideration requires the numerical solution of the coupled electrohydrodynamic equations on a discrete lattice. We emphasize that the high numerical complexity of this scheme is expected to shadow the physical transparency of our simpler theory. Then, the Stokes equation was solved with the no-slip boundary condition. Future works may consider the effect of a finite slip length on the translocation process~\cite{Boc1,Boc2,Boc3}. Furthermore, the electrostatic 1l formalism neglects the formation of ionic pairs between monovalent anions and poyvalent cations~\cite{Pianegonda2005,Diehl2006,dosSantos2010}. We note that despite this limitation, the 1l theory has been shown to agree with the MC simulations of polyvalent solutions in charged cylindrical nanopores~\cite{Buyuk2014}. Moreover, our solvent-implicit electrolyte model does not account for the solvent charge structure. It should be however noted that due to the large radius of the solid state pores considered in our work, interfacial effects associated with the solvent charge structure are not expected to affect qualitatively our physical conclusions~\cite{Buyuk2014II}. In addition, our rigid polyelectrolyte model neglects the entropic polymer conformations. Within the unified theory of ionic and polymer fluctuations developed by Tsonchev et al.~\cite{Duncan1999}, the polymer flexibility can be incorporated into our model but this tremendous task is beyond the scope of our article.  In our model, we also neglected the variations of the surface charges with the salt density.  We are currently working on the incorporation of the pH-controlled charge regulation mechanism into the theory. The gradual improvement of our model upon these extensions will enable a more extensive confrontation with ion and polymer conductivity experiments. We finally note that the inverted EO flow-assisted polymer transport mechanism can be easily corroborated by standard polymer translocation experiments involving anionic membrane nanopores.

\smallskip
\appendix
\section{Computation of the 1l-level electrostatic potential $\phi(r)$}
\label{avpot}
We explain here the computation of the correlation-corrected average electrostatic potential $\phi(r)$ required for the calculation of the drift velocity in Eq. (10) of the main text. The underlying 1l formalism being valid for dielectrically continuous media, the potential $\phi(r)$ will be computed by neglecting the dielectric jumps in the system.  According to the 1l-theory of charge correlations~\cite{Buyuk2014,the16}, the average potential in the pore is given by
\be\label{16}
\phi(r)=\phi_0(r)+\phi_c(r).
\ee
In Eq.~(\ref{16}), the MF  component $\phi_0(r)$ solves the radial PB equation
\bea\label{17}
&&\frac{1}{4\pi\ell_Br}\partial_r\left[r\partial_r\phi_0(r)\right]+\sum_{i=1}^pq_in_i(r)\nonumber\\
&&=-\sigma_m\delta(r-d)-\sigma_p\delta(r-a),
\eea
with the ionic number density function
\be\label{18}
n_i(r)=\rho_{bi}e^{-q_i\phi_0(r)}\theta(d-r)\theta(r-a).
\ee
Together with the Gauss' laws $\phi_0'(a^+)=-4\pi\ell_B\sigma_p$ and $\phi_0'(d^-)=4\pi\ell_B\sigma_m$, Eq.~(\ref{17}) can be easily solved by numerical discretization. Then, the potential component $\phi_c(r)$ associated with charge correlations reads
\be
\label{19}
\phi_c(r)=\int_a^d\mathrm{d}r'r'\tG_{0}(r,r';k=0)\delta\sigma(r').
\ee
Eq.~(\ref{19}) includes the Fourier-transform of the electrostatic Green's function $G(\br,\br')$ defined by
\be
\label{20}
G(\br,\br')=\sum_{n=-\infty}^{\infty}e^{in(\theta-\theta')}\int_{-\infty}^\infty\frac{\mathrm{d}k}{4\pi^2}\tG_n(r,r;k)e^{ik(z-z')},
\ee
with $\theta$ the polar angle in the $x-y$ plane. The Green's function is the solution of the kernel equation
\be\label{21}
\left[\nabla^2-\chi^2(r)\right]G(\br,\br')=-4\pi\ell_B\delta(\br-\br'),
\ee
with the screening  function $\chi^2(r)=4\pi\ell_B\sum_iq_i^2n_i(r)$.  Eq.~(\ref{19}) also contains the charge density excess
\be\label{22}
\delta\sigma(r)=-\frac{1}{2}\sum_{i=1}^pq_i^3n_i(r)\delta G(r),
\ee
with the ionic self-energy corresponding to the renormalized equal-point correlation function 
\be
\label{23}
\delta G(r)=\sum_{n=-\infty}^{\infty}\int_0^\infty\frac{\mathrm{d}k}{2\pi^2}\left[\tG_n(r,r;k)-\tG_{bn}(r,r;k)\right],
\ee
where the Fourier-transformed bulk Green's function is
\be\label{27}
\tG_{bn}(r,r';k)=4\pi\ell_B\mathrm{I}_n(p_br_<)\mathrm{K}_n(p_br_>).
\ee
In Eq.~(\ref{27}), we used the radial variables
\be\label{27II}
r_<=\mathrm{min}(r,r')\;;\hspace{5mm}r_>=\mathrm{max}(r,r').
\ee

According to Eqs.~(\ref{19}) and~(\ref{22}), the computation of the correction term $\phi_c(r)$ necessitates the knowledge of the Green's function $G(\br,\br')$ solving Eq.~(\ref{21}). In order to solve this kernel equation, we exploit the cylindrical symmetry and insert the Fourier expansion~(\ref{20}) into Eq.~(\ref{21}). Then, we use the definition of the Green's function $\int\mathrm{d}\br''G^{-1}(\br,\br'')G(\br'',\br')=\delta(\br-\br')$ to recast Eq.~(\ref{21}) as an integral relation, 
\bea
\label{24}
\tG_n(r,r';k)&=&\tG^{(0)}_n(r,r';k)\\
&&+\int_a^d\mathrm{d}uu\tG^{(0)}_n(r,u;k)\delta n(u)\tG_n(u,r';k).\nonumber
\eea
Eq.~(\ref{24}) involves  the excess ion density function
\be\label{25}
\delta n(r)=\sum_{i=1}^p\rho_{bi}q_i^2\left[1-e^{-q_i\phi_0(r)}\right]
\ee
and the reference potential $\tG^{(0)}_n(r,r';k)$. The latter corresponds to the Fourier-transform of the DH Green's function~\cite{Buyuk2014,the16}
\be
\label{26}
\tG^{(0)}_n(r,r';k)=\tG_{bn}(r,r';k)+\delta\tG^{(0)}_n(r,r';k),
\ee
where the inhomogeneous part reads
\bea
\label{28}
\delta\tG^{(0)}_n(r,r';k)&=&\frac{4\pi\ell_B}{g_1g_2-1}\left\{g_1\mathrm{I}_n(p_br_<)\mathrm{I}_n(p_br_>)\right.\\
&&\hspace{1.5cm}+g_2\mathrm{K}_n(p_br_<)\mathrm{K}_n(p_br_>)\nonumber\\
&&\hspace{1.5cm}+\mathrm{I}_n(p_br_<)\mathrm{K}_n(p_br_>)\nonumber\\
&&\hspace{1.5cm}\left.+\mathrm{K}_n(p_br_<)\mathrm{I}_n(p_br_>)\right\}.\nonumber
\eea
In Eq.~(\ref{28}), we introduced the geometric factors
\bea
\label{29}
g_1&=&\frac{\mathrm{I}_n(ka)\mathrm{K}'_n(p_ba)-\mathrm{I}'_n(ka)\mathrm{K}_n(p_ba)}{\mathrm{I}'_n(ka)\mathrm{I}_n(p_ba)-\mathrm{I}_n(ka)\mathrm{I}'_n(p_ba)},\\
\label{30}
g_2&=&\frac{\mathrm{K}_n(kd)\mathrm{I}'_n(p_bd)-\mathrm{K}'_n(kd)\mathrm{I}_n(p_bd)}{\mathrm{K}'_n(kd)\mathrm{K}_n(p_bd)-\mathrm{K}_n(kd)\mathrm{K}'_n(p_bd)}
\eea
accounting for the presence of the concentric nanopore and the DNA molecule. 

After obtaining the MF potential $\phi_0(r)$ in Eq.~(\ref{25}) from the solution of Eq.~(\ref{17}), the integral Eq.~(\ref{24}) can be numerically solved by iteration. The details of this iterative scheme can be found in Ref.~\cite{Buyuk2014}. The resulting Green's function $\tG_n(r,r';k)$ is to be used next in Eqs.~(\ref{19}) and~(\ref{22})-(\ref{23}) in order to obtain the potential correction $\phi_c(r)$. The substitution of the 1l potential~(\ref{16}) into Eq. (10) of the main text provides us with the correlation-corrected drift velocity. We finally note that in Eq. (29) of the main text, the liquid charge density is defined as 
\be\label{chd}
\rho_{c}(r)=\sum_{i=1}^pq_i\rho_{i}(r), 
\ee
with the correlation-corrected ionic number density
\be\label{den1l}
\rho_i(r)=\rho_{ib}e^{-q_i\phi_0(r)}\left[1-q_i\phi_c(r)-\frac{q_i^2}{2} \delta G(r)\right].
\ee
\\
\acknowledgements  This work was performed as part of the Academy of Finland Centre of Excellence program (project 312298).


\begin{thebibliography}{99}
\bibitem {rem1} In this equality, the coordinate $z_p$ should be considered as an adiabatic variable.
\bibitem {rem2} By imposing the absorbing boundary condition at the pore exit, we assume that the polymer leaving the pore at $z_p=L_m+L_p$ is immediately removed from the system.
\bibitem {rem3} In our model, the electric field $\bE$ is oriented along the negative $z$ axis, i.e. oppositely to the field direction of Ref.~\cite{e21}. Therefore, in Fig.~\ref{fig2}(a), the sign of the experimental DNA mobility data obtained from Fig.2 of Ref.~\cite{e21} has been inverted.
\end{thebibliography}
\end{document}